\documentclass[cits]{PoS}

\title{Pion transition form factor in the Regge approach}

\ShortTitle{Pion transition form factor in the Regge approach}

\author{\speaker{Wojciech Broniowski}%
         \thanks{Supported by Polish Ministry of Science
                 and Higher Education, grants N~N202~263438 and N~N202~249235}\\
        The H. Niewodnicza\'nski Institute of Nuclear Physics, Polish Academy of Sciences, PL-31342~Krak\'ow, Poland\\
        Institute of Physics, Jan Kochanowski University, PL-25406~Kielce, Poland\\
        E-mail: \email{Wojciech.Broniowski@ifj.edu.pl}}

\author{Enrique Ruiz Arriola%
         \thanks{Supported by Spanish DGI and
         FEDER funds with grant FIS2005-00810, Junta de Andaluc{\'\i}a
         grant FQM225-05, and EU Integrated Infrastructure Initiative
         Hadron Physics Project contract RII3-CT-2004-506078}\\
        Departamento de F\'{\i}sica At\'omica, Molecular y Nuclear, Universidad de Granada, E-18071~Granada, Spain\\
        E-mail: \email{earriola@ugr.es}}

\abstract{We explore the BaBar puzzle within the Regge approach. After reviewing the 
 chiral quark models in applications to PDF and PDA of the pion, we argue
 that variants of these models, fulfilling the chiral
 anomaly, may in fact violate the second Terazawa-West unitarity bound, which
 is based on unverified assumptions for the real part of the
 amplitude. Consequently, the transition form factor need not
 vanish at large values of the photon virtuality. Then we show that the
 experimental data may be properly explained with incomplete
 vector-meson dominance in a simple model with one state, as well as
 in more sophisticated radial Regge models including infinitely many states. We
 also consider the experimental constraint from the rare $Z \to \pi_0
 \gamma$ decay, which is comfortably satisfied in our approach. 
 Finally, we point out that the photon momentum asymmetry parameter may
 noticeably influence the precision fits to the data.}

\FullConference{Light Cone 2010 - LC2010\\
		June 14-18, 2010\\
		Valencia, Spain}

\def\slashchar#1{\setbox0=\hbox{$#1$}
   \dimen0=\wd0 \setbox1=\hbox{/} \dimen1=\wd1
   \ifdim\dimen0>\dimen1 \rlap{\hbox to \dimen0{\hfil/\hfil}} #1
   \else  \rlap{\hbox to \dimen1{\hfil$#1$\hfil}} / \fi}

\begin{document}

\section{Introduction}

After the release of the BaBar data \cite{Aubert:2009mc} for the
pion-photon transition form factor, $F_{\pi^0 \gamma \gamma^\ast}$,
our community is in deep shock, as the conventional approach to the
gold-plated exclusive process at very high Euclidean momenta $Q$,
based on 1.~factorization and 2.~(leading-twist) pQCD
evolution~\cite{Efremov:1979qk,Lepage:1979zb,Lepage:1980fj,Brodsky:1981rp,delAguila:1981nk,Braaten:1982yp,Kadantseva:1985kb,%
  Bakulev:2001pa,Bakulev:2002uc,Bakulev:2003cs,Mikhailov:2009kf},
seems to be invalid~\cite{Isgur:1984jm,Isgur:1988iw}.  Indeed, the
quantity $Q^2 F_{\pi^0 \gamma \gamma^\ast}(Q^2)$ goes visibly above
the famous Brodsky-Lepage limit,
\begin{eqnarray}
Q^2 F_{\pi^0 \gamma \gamma^\ast}(Q^2) \to \frac{2 f_\pi}{N_c}\int_0^1 dx \frac {\phi_{\rm as}(x)}{x} 
=  \frac{6 f_\pi}{N_c} = 2 f_\pi, \label{BL}
\end{eqnarray}
at momenta $Q^2 > 15~{\rm GeV}^2$.

Several ideas, abandoning assumptions 1. and 2., have been proposed to
solve the ``BaBar problem''. Radyushkin~\cite{Radyushkin:2009zg} and
Polyakov~\cite{Polyakov:2009je} advocated that the possible
nonvanishing of the PDA at the end points (as found by the present
authors in chiral quark models at the low-energy quark-model scale
\cite{RuizArriola:2002bp}), together with essentially switched-off
evolution and regulated quark propagators, is capable of reproducing
the data in the CLEO and BaBar domain. In this approach $F_{\pi^0
  \gamma \gamma^\ast} \sim \log(Q^2/\mu^2)/Q^2$, with the $\log$
indicating the breaking of factorization (note that the same
asymptotics follows in the Spectral Quark Model (SQM), cf. Eq.~(14.1)
of~\cite{RuizArriola:2003bs}). Dorokhov~\cite{Dorokhov:2009dg,Dorokhov:2009zx,Dorokhov:2009jd}
proposed the use of the fixed-mass chiral quark model to evaluate the
triangle diagram of Fig.~\ref{fig:tri}, which is capable of
reproducing the data at high $Q^2$, however the fitted value of the
constituent quark mass is very low, $M \sim 135$~MeV. In this the form
$F_{\pi^0 \gamma \gamma^\ast} \sim [\log(Q^2/\mu^2)]^2/Q^2$.  A
possible need of higher-twist terms has been brought up
in~\cite{Noguera:2010fe}.  The calculation of~\cite{Kotko:2009ij} in
the nonlocal chiral quark model inspired by the instanton-liquid model
of QCD produced the result in agreement with the data at lower values
of $Q^2$ and complying to the limit (\ref{BL}). On the other hand,
Dorokhov~\cite{Dorokhov:2010bz} considered a modified pion-meson
vertex \cite{Holdom:1990iq} in the nonlocal model and found agreement
in the whole available data range, albeit again with a very low
constituent quark mass. Asymptotically, in this approach $F_{\pi^0
  \gamma \gamma^\ast}\sim \log(Q^2/\mu^2)/Q^2$.
The influence of nonperturbative gluonic components of the
pion has been considered in~\cite{Kochelev:2009nz}. Concerns on the validity of factorization
were discussed in~\cite{Mikhailov:2009sa}. It was also shown that the 
pronounced growth of $Q^2 F_{\pi^0 \gamma \gamma^\ast}$ between 10 and 40 GeV$^2$ cannot be explained via
higher-order pQCD corrections at the NNLO level~\cite{Mikhailov:2009kf}.
Further developments may be found in~\cite{Chernyak:2009dj,Khodjamirian:2009ib,Li:2009pr,Wu:2010zc}

This talk is based on~\cite{Arriola:2010aq}.

\section{General constraints}

\begin{figure}[b]
\begin{center}
\includegraphics[width=0.375\textwidth]{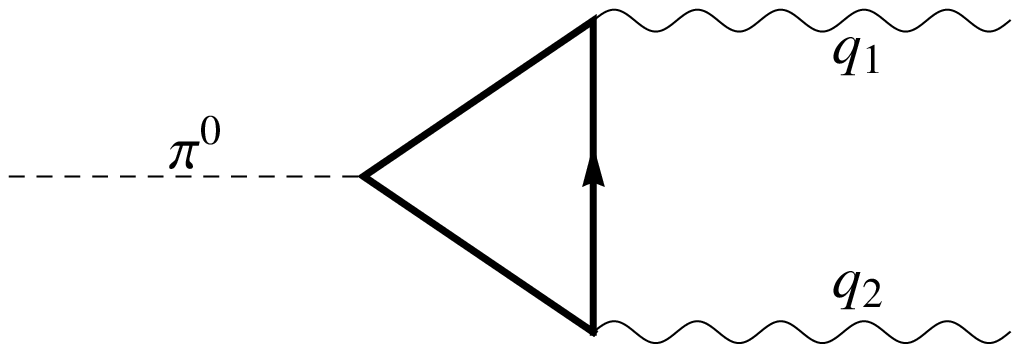} ~~~ \includegraphics[width=0.375\textwidth]{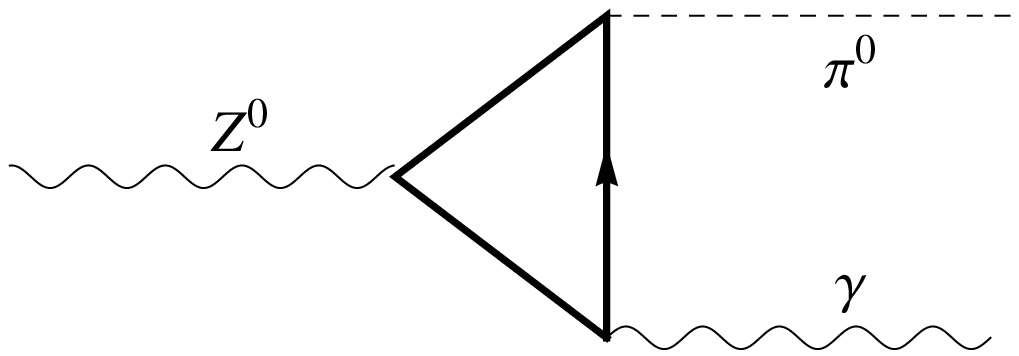} 
\end{center}
\caption{The quark-model diagrams used to evaluate the pion transition form factor (left)
 and the $Z_0 \to \pi^0 \gamma$ decay (right). The crossed diagrams not shown.
\label{fig:tri}}
\end{figure}

Consider the process of the left part of Fig.~\ref{fig:tri} with the general kinematics
\begin{eqnarray} 
q_1^2=-\frac{1+A}{2} Q^2, \; q_2^2=-\frac{1-A}{2} Q^2, \;\; -1 \le A \le 1. \label{kin}
\end{eqnarray}
The chiral anomaly~\cite{Adler:1969gk,Bell:1969ts} fixes $F_{\pi^0 \gamma^\ast \gamma^\ast} (Q^2=0,A)= \frac{1}{4\pi^2 f_\pi}$.

On the other, the high-$Q^2$ behavior of $F_{\pi^0 \gamma
  \gamma^\ast}$ is formally limited by the Terazawa-West
(TW)~\cite{Terazawa:1973hk,West:1973gd,Terazawa:1973tb} unitarity
bounds, recently brought up by Dorokhov~\cite{Dorokhov:2009jd} and
further elaborated in~\cite{Arriola:2010aq}.  The derivation of the TW
bounds uses the Schwarz inequality involving sums of the matrix
elements $\langle 0 | J_\mu(0) | n \rangle $ and $\langle \pi^a(q) |
J_\mu(0) | n \rangle $, entering the vacuum polarization and the
parton distribution function (PDF) of the pion. Schematically, one has
$| \langle \pi | J J |0 \rangle | \le \left ( |\langle 0 | J J | 0
\rangle ||\langle \pi | J J | \pi \rangle | \right )^{1/2}$.  Then one
derives the first bound,
\begin{eqnarray}
{\rm Im} F_{\pi^0 \gamma \gamma^\ast}(q^2) = {\cal O}(1/\sqrt{q^2}) \;\;\;\; ({\rm TW~I})
\end{eqnarray}
holding at {\em time-like} momenta, $q^2>4 m_\pi^2$. 
If there are {\em no polynomial terms} in the real part of $F_{\pi^0 \gamma \gamma^\ast}$ (which is an {\em assumption}), then
$\left | F_{\pi^0 \gamma \gamma^\ast}(q^2) \right | = {\cal O}(1/\sqrt{q^2})$. 
A dispersion relation yields~\cite{Terazawa:1973tb} the second bound, 
\begin{eqnarray}
{\left | F_{\pi^0 \gamma \gamma^\ast}(Q^2) \right | = {\cal O}(1/Q) \;\;\;\; ({\rm TW~II})}
\end{eqnarray}
valid for all momenta, also large {\em space-like} momenta $Q$.
The constant in the bound may be explicitly given~\cite{Terazawa:1973tb} in terms of the photon spectral density and the
pion structure function,
\begin{eqnarray}
\left | F_{\pi^0 \gamma \gamma^\ast}(Q^2) \right | < \frac{2 \sqrt{\Pi(\infty)}}{Q} 
\int_0^1 dx \sqrt{\frac{F_1(x,Q^2)}{x(1-x)}} 
\end{eqnarray}
where $\Pi(s) =s/(16\pi^3 \alpha_{\rm QED}^2) \sigma_{e^+ e^- \to {\rm hadrons}}(s)$ and $\Pi(\infty) = 1/({12\pi^2}) \sum_i e_i^2$, with 
$e_i$ denting the quark charges.
With the SMRS~\cite{Sutton:1991ay} and GRV~\cite{Gluck:1999xe}
parameterizations for $F_1$ we obtain (for $Q^2$ in the range
$10-40~{\rm GeV}^2$)
\begin{eqnarray}
|F_{\pi^0 \gamma \gamma^\ast}(Q^2)| < \frac{0.85(1)}{Q} \quad {\rm (LO),}  \;\;\;\;\; < \frac{0.75(1)}{Q} \quad {\rm (NLO),} \label{TW2}
\end{eqnarray}
where LO and NLO refer to Leading Order and Next-to-Leading Order in
the QCD evolution, respectively.  Thus the TW II bound is
``inefficient'' in the present experimental range, as it goes an order
of magnitude above the BaBar data. 

Finally, another interesting experimental bound comes from the rare $Z \to \pi^0 \gamma$
decay~\cite{Jacob:1989pw}, which probes the time-like value
$q^2=M_Z^2$. The process is shown in the right part of Fig.~\ref{fig:tri}. 
Since only the vector coupling of the $Z^0$ boson to the quark contributes,
\begin{eqnarray}
\frac{F_{Z \to \pi^0 \gamma} (q^2)}{F_{Z \to \pi^0 \gamma} (0)}=
\frac{F_{\pi^0 \gamma^\ast \gamma} (q^2)}{F_{\pi_0 \gamma^\ast \gamma} (0)}.
\end{eqnarray}
The experimental limit given by the Particle Data Group~\cite{Amsler:2008zzb}, $\Gamma( Z^0 \to \pi^0 \gamma) < 5
\times 10^ {-5} \Gamma_{\rm tot} (Z^0)= 10.25 \times 10^{-5} {\rm GeV}$, implies  
\begin{eqnarray}
{|F_{Z^0 \to \pi^0 \gamma} (M_Z^2)/F_{Z^0 \to \pi^0 \gamma} (0)| < 0.17}. \label{z:b}
\end{eqnarray}

\section{Subtracted dispersion relation and violation of TW II}

The assumption of the absence of the polynomial terms, necessary for
TW II to hold, is equivalent to the validity of the unsubtracted
dispersion relation for $F_{\pi^0 \gamma^\ast \gamma}$.  Clearly, pQCD
with factorization leads to $F_{\pi^0 \gamma^\ast \gamma}$ vanishing
at $Q \to \infty$, however, since now these assumptions are questioned
(see the Introduction), one may consider the situation where
subtraction constants are necessary in the dispersion relation for
$F_{\pi^0 \gamma^\ast \gamma}$ \cite{Arriola:2010aq}.  Note that even
if the form factor vanishes at infinity, one can write a subtracted
relation
\begin{eqnarray}
F(t) - F(0) = \frac1\pi \int_{s_0}^{4
  \Lambda^2}\ \frac{t}{s}\frac{{\rm Im} F (s)}{s-t} ds + \frac1\pi
\int_{4 \Lambda^2}^\infty\ \frac{t}{s}\frac{{\rm Im} F(s)}{s-t} ds.
\label{eq:DR}
\end{eqnarray}
If $\Lambda$ is large ($\Lambda^2>Q^2$), the second term is very
slowly varying with $Q^2$ and mimics a constant.  In particular, for
${\rm Im} F (s) \sim 1/\sqrt{s}$ it behaves as $1/\Lambda + {\cal
  O}(1/Q)$.  Thus the appearance of the constant term need not be
taken as a fundamental problem, as it may represent the unknown
high-energy data.  We will analyze below a quantitative lower bound
for a possible high energy mass scale. 

The bottom line is that one may well consider the case where $F_{\pi^0
  \gamma^\ast \gamma}(Q^2)$ does not vanish at $Q^2 \to \infty$.
Below we will provide an explicit field-theoretic example where this
is the case (the Georgi-Manohar (GM) model~\cite{Manohar:1983md}).
Motivated by this we will also incorporate a constant term in the fits
of the BaBar data.

\section{Mini-review of chiral quark models}

Before discussing the GM model let us briefly review the chiral quark
models and, in particular, their predictions for the {\em soft} matrix
elements entering the (factorized) high-energy processes. The purpose
is to convince the reader that these models, based on {\em chiral
  symmetry breaking} as the key dynamical ingredient, are very
successful for numerous processes involving the pions and photons.
Chiral quark models are cast in a covariant Lagrangian form and in
general exhibit no factorization. They carry relatively few
parameters, traded for $f_\pi$, $m_\pi$, \dots The large-$N_c$ limit
is implicit, however, confinement is absent and one needs to be
careful not to open the $q \overline{q}$ production threshold. This
limits the applicability range, however, no problems arise with
space-like external momenta, as in the present analysis of the BaBar
puzzle.  Finally, predictions of chiral quark models hold at a
low-energy {\em quark-model} scale \cite{Broniowski:2007si} and {QCD
  evolution} is necessary to reach higher scales of experiments or
lattice simulations.

\begin{figure}[tb]
\begin{center}
\includegraphics[width=7.5cm]{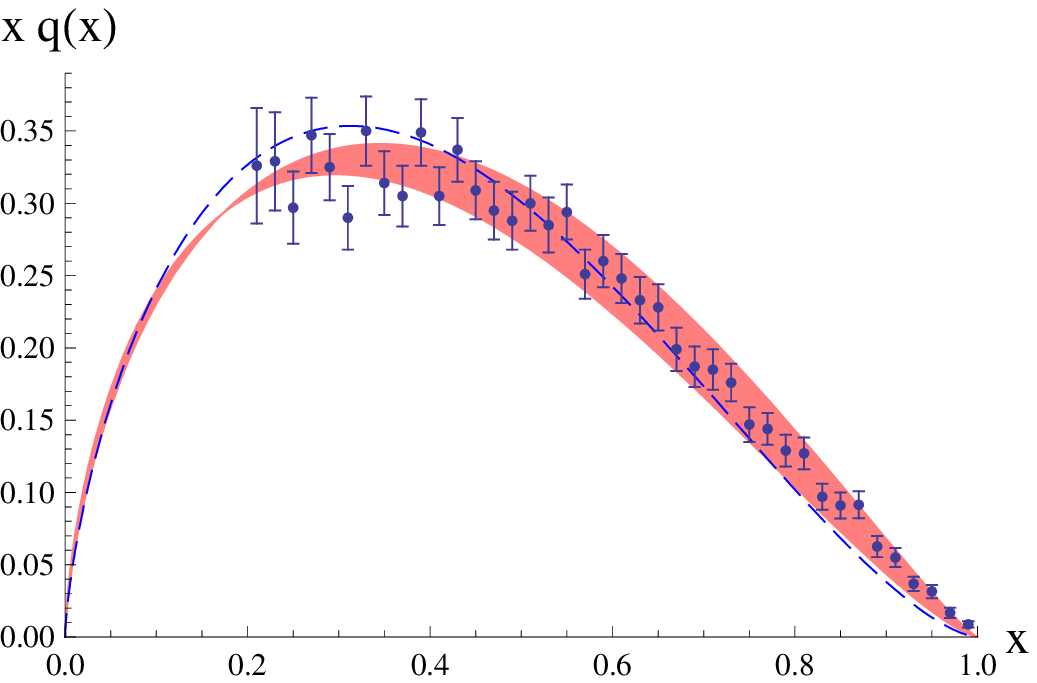} 
\hfill \includegraphics[width=7.5cm]{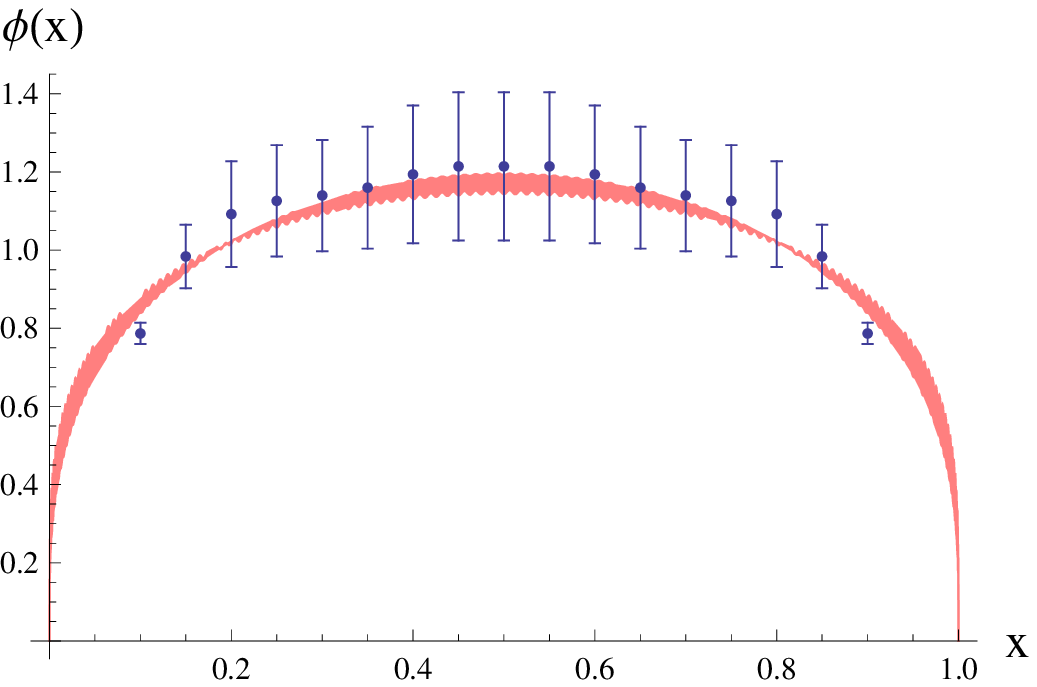} 
\end{center}
\caption{Left: The NJL prediction for the valence PDF of the pion
  evolved to the experimental scale $\mu=4$~GeV (band).  The data
  points come from the analysis of the Drell-Yan data from the E615
  experiment \cite{Conway:1989fs}. The dashed line shows the
  reanalysis of the data from~\cite{Wijesooriya:2005ir}.  Right: The
  NJL prediction for the PDA of the pion evolved to the lattice scale
  $\mu=0.5 {\rm GeV} $ (band) compared to the transverse lattice
  data~\cite{Dalley:2002nj}. The width of the bands indicates the
  uncertainty in the quark-model scale $\mu_0$.
\label{fig:qmp}}
\end{figure}

We start with the valence PDF of the pion, where the NJL model gives
\cite{Davidson:1994uv}
\begin{eqnarray} 
q(x)=1 \label{pdf:1}
\end{eqnarray}
(note that the same {\em constant} PDF for two constituents follows
from AdS/CFT approach of \cite{Brodsky:2007hb}).  Arguments based on
the momentum sum rule \cite{Broniowski:2007si} determine the
renormalization scale where Eq.~(\ref{pdf:1}) holds: $\mu_0 \sim
320$~MeV. Then at the scale $\mu=2$~GeV the valence quarks carry 47\%
of the momentum of the pion, as requested by experimentally motivated 
parameterizations.  The value of the coupling constant at the
quark-model scale is $\alpha(\mu_0)/\pi= 0.68$.  After the evolution
the NJL prediction agrees remarkably well with the experimental data
(left panel of Fig.~\ref{fig:qmp}).

In the NJL model, the PDA of the pion is also constant at the quark-model scale \cite{RuizArriola:2002bp}, 
\begin{eqnarray}
\phi(x)=1 \label{pda}
\end{eqnarray}
(note that this result is different from the AdS/CFT prediction
of~\cite{Brodsky:2007hb}, $\phi(x) \sim \sqrt{x(1-x)}$, it is also far
from the asymptotic form $6x(1-x)$). The PDA~(\ref{pda}) does not
vanish at the end points, which is the focal point
of~\cite{Radyushkin:2009zg,Polyakov:2009je}. However, the LO ERBL
evolution makes $\phi(x)$ {vanish} at the end-points, with the form
$\phi(x) \sim x^{2C_F/\beta_0 \, \log[\alpha(\mu_0)/\alpha(\mu)]}$
near $x=0$ \cite{Efremov:1979qk,Broniowski:2007si}.  Thus maintaining the
nonvanishing at the end points requires switching off the QCD
evolution~\cite{Radyushkin:2009zg}.

Finally, we mention the NJL results for the gravitational form factor of the pion,
discussed in detail in~\cite{Broniowski:2008hx}. The quark-model
relation holds for the radii related to the spin-2 gravitational form
factor and the charge form factor, $\langle r^2 \rangle_\Theta =
\frac{1}{2}\langle r^2 \rangle_V$.  Therefore matter is more
concentrated than charge, in agreement with the recent AdS/CFT
results~\cite{Brodsky:2008pf}.  Note that there is no contradiction
between having a constant vertex function and a finite pion size. This
issue is tightly linked with the fulfillment of the electromagnetic and
chiral Ward identities, as discussed, e.g., in~\cite{RuizArriola:2002wr}.

\section{Violating TW II}

We now recall a model which reproduces the results of the previous
section, but violates TW~II: the Georgi-Manohar (GM)
model~\cite{Manohar:1983md}. It is obtained from the NJL model by
carrying out the chiral rotation (which is innocuous) and then
introducing $g_A$ of the quark which may be different from
unity~\cite{Broniowski:1993nb}. The Lagrangian of the GM model is
\begin{eqnarray}
 L=\bar q \left ( i \slashchar{\partial} + g_A^Q \slashchar{A} \gamma_5  - M   \right ) q 
+ \frac{f^2}{4} {\rm Tr} \left ( \partial_\mu U^\dagger \partial^\mu U \right ) + {\rm WZW},
\end{eqnarray}
where WZW stands for the Wess-Zumino-Witten term, while 
\begin{eqnarray} 
A_\mu=\frac{i}{2}(u^\dagger \partial_\mu u-u \partial_\mu u^\dagger), \;\; u=e^{i \vec{\pi}\cdot\vec{\tau} /(2f)},\;\; U=u^2. 
\end{eqnarray}
The resulting pion-photon transition form factor is
\begin{eqnarray}
F_{\pi^0 \gamma \gamma^*} (Q^2) = \frac{1}{4 \pi^2 f_\pi} + \frac{g_A^Q}{4
  \pi^2 f_\pi} \left[ G(Q^2)-1 \right], \;\; G(Q^2) = \frac{2 M^2}{Q^2} \int_0^1 \frac{dx }{x} \log\left[1+x(1-x)
  \frac{Q^2}{M^2} \right]
\end{eqnarray}
As we can see, the anomaly is satisfied, but for $g_A^Q \neq 1$ {\em
  no vanishing} at $Q^2 \to \infty$ occurs,
\begin{eqnarray}
F_{\pi^0 \gamma \gamma^*} (Q^2) = \frac{1-g_A^Q}{4 \pi^2 f_\pi}  + 
\frac{g_A^Q M^2}{4 \pi^2 f_\pi} \frac{\left[\log (Q^2/M^2)\right]^2}{Q^2} + \dots 
\end{eqnarray}
In the Spectral Quark Model \cite{RuizArriola:2003bs} one has instead 
\begin{eqnarray}
G(Q^2)= \frac{1}{3}\left[ \frac{2 m_\rho^2}{m_\rho^2+Q^2}+\frac{m_\rho^2}{Q^2} \log
\frac{m_\rho^2+Q^2}{m_\rho} \right], \; F_{\pi^0 \gamma \gamma^*} = \frac{1-g_A^Q}{4 \pi^2 f_\pi} +
\frac{g_A^Q m_\rho^2}{12 \pi^2 f_\pi} \frac{\log
    \frac{Q^2}{m_\rho^2}}{Q^2} + \dots 
\end{eqnarray}
With $g_A^Q = 1 $ this model fulfills the result of~\cite{Radyushkin:2009zg}
with the mass scale $m_\rho$.

Unfortunately, precise chiral-quark-model fits of $F_{\pi^0 \gamma
  \gamma^*}$ based on above formulas in the {\em whole} $Q^2$ range
are not satisfactory, so in the following we proceed to more general
analyses based on Incomplete Vector Meson Dominance
(IVMD)~\cite{Knecht:2001xc,Nyffeler:2009uw}.

\section{Incomplete VMD and Regge models}

As shown in the previous Section, it is possible to formulate a
field-theoretic model consistent with all formal requirements which
violates TW~II, leading to an asymptotically constant $F_{\pi^0 \gamma
  \gamma^*}(Q^2)$. With this finding in mind we explore the idea of
IVMD and the Regge models. Regge models with (infinitely many)
tree-level meson and glueball exchanges are a realization of the
large-$N_c$ limit which, unlike the current quark models, incorporate
confinement and comply with quark-hadron duality by generating
meromorphic two-point functions for color singlet currents.

\begin{figure}[tb]
\includegraphics[width=0.49\textwidth]{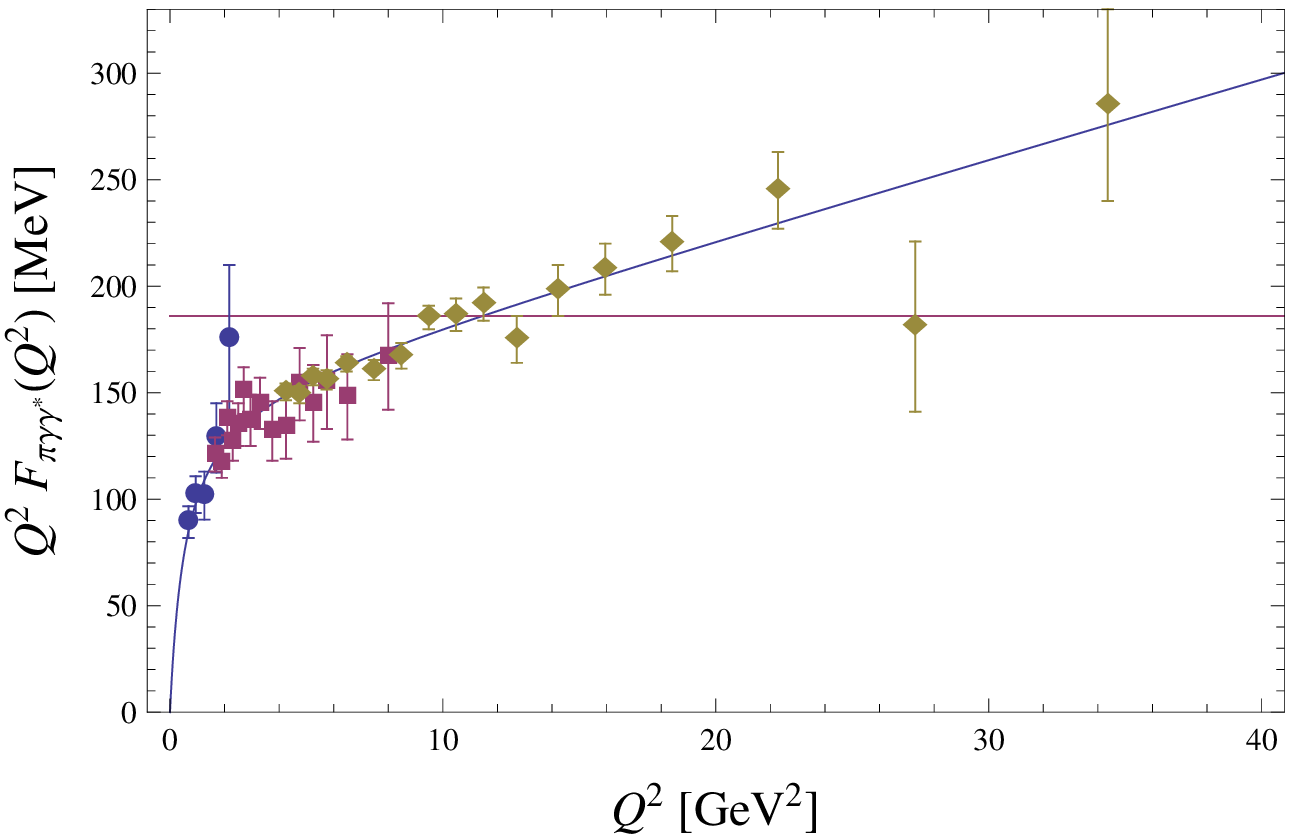} \hfill \includegraphics[width=0.49\textwidth]{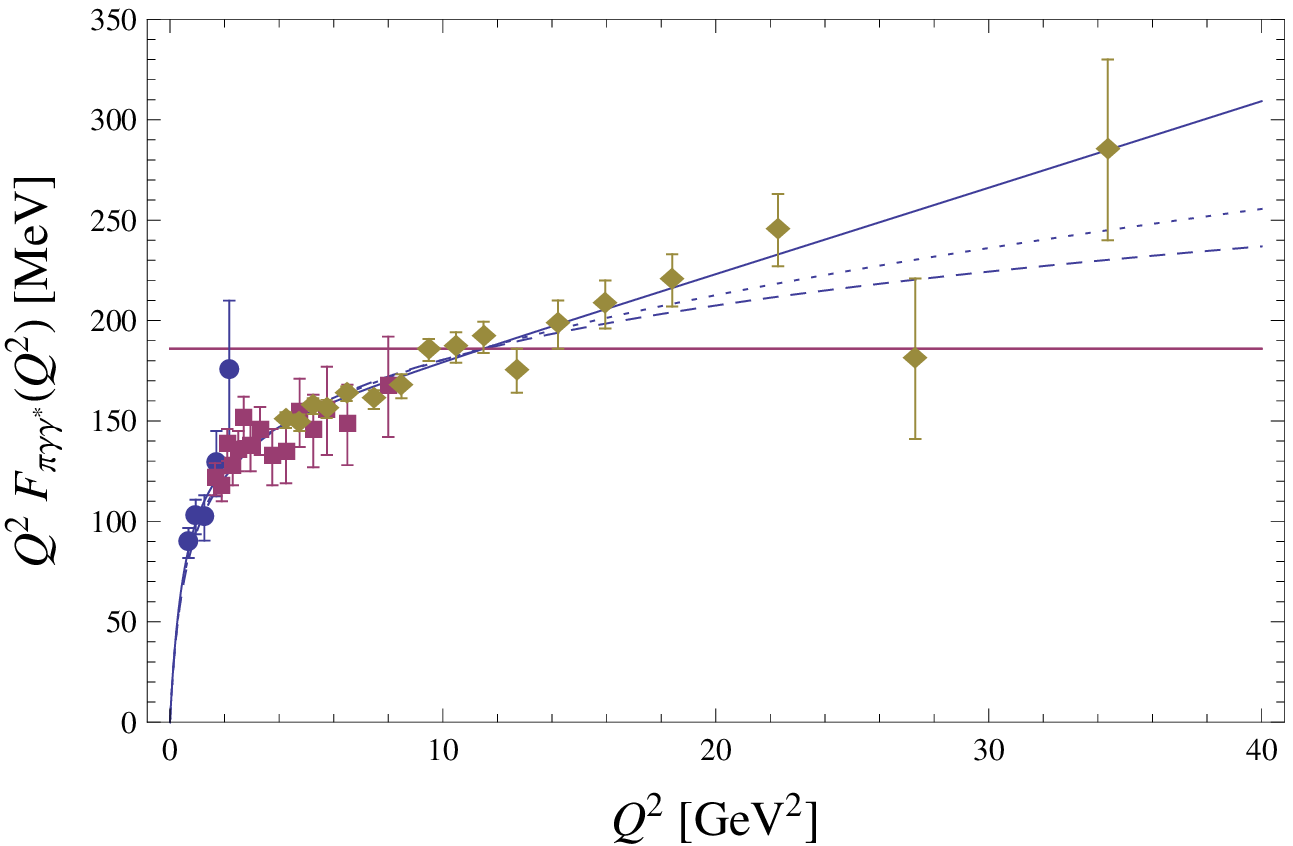} 
\caption{Left: predictions of the one-state IVMD model (line). Right: 
predictions of various Regge models: the Veneziano-Dominguez model $b=1.81$ (dashed line),
model with the first pole separated and fixed $b=1.5$ (dotted line), and the
subtracted Regge model (solid line).
The dots, squares, and diamonds correspond to the 
CELLO~\cite{Behrend:1990sr}, CLEO~\cite{Gronberg:1997fj}, and BaBar~\cite{Aubert:2009mc} data, correspondingly. 
\label{fig:one}}
\end{figure}

The simplest IVMD model which comes to mind has just one state
saturating the subtracted dispersion relation for $F_{\pi^0 \gamma
  \gamma^*}$. Then
\begin{eqnarray}
F_{\pi^0 \gamma \gamma^*} (Q^2) = \frac{1}{4 \pi^2 f_\pi} \left[ 1 -c \frac{Q^2}{M_V^2+Q^2} \right].
\label{eq:IVMD}
\end{eqnarray}
The fit to the combined CELLO~\cite{Behrend:1990sr},
CLEO~\cite{Gronberg:1997fj}, and BaBar~\cite{Aubert:2009mc} data yields
\mbox{${c=0.986(2)}$}, {$M_V=748(14)~{\rm MeV}$}, with $\chi^2/{\rm DOF}=0.7$,
thus $(1-c)$ is significantly different from 0 although numerically
small. On the other hand, the fit to the CLEO data only gives $c =
0.998(18)$, $M_V=777(44)~{\rm MeV}$, $\chi^2/{\rm DOF}=0.54$, with
$(1-c)$ compatible with 0, or complete VMD.  The result of the fit to
the combined data is shown in the left panel of Fig.~\ref{fig:one},
displaying a remarkable agreement in the whole experimentally available
range.  The radius squared
\begin{eqnarray} 
b_\pi = -\left[\frac1{F_{\pi^0 \gamma \gamma^\ast} (Q)}
\frac{d}{d Q^2} F_{\pi^0 \gamma \gamma^\ast} (Q) \right]\Big|_{Q^2=0}
\end{eqnarray}
becomes $b_\pi = \frac{c}{M_V^2} = 1.76(7)~{\rm GeV}^{-2}$, compared
to the PDG value \mbox{$b_\pi = (1.76 \pm 0.22 ) {\rm GeV}^{-2}$}.
The constrain from the rare $Z^0$ decay is satisfied comfortably, as
$|F_{Z \to \pi^0 \gamma} (M_Z^2)/F_{Z \to \pi^0 \gamma} (0)| =
0.014(2)$, an order of magnitude less than $0.17$ of Eq.~(\ref{z:b}).

Next, we consider radial Regge models, with $M_n^2 = M_V^2 + a n$,
recalling that the large-$N_c$ QCD involves tree-level diagrams with
infinitely many states, including the radial excitations.  A
particular realization, the Veneziano-Dominguez
model~\cite{Veneziano:1974dr,Dominguez:1983aa}, allows to control the
asymptotic behavior of the form factor with a single parameter, $b$:
\begin{eqnarray}
&& F_{\pi^0 \gamma \gamma^*}(t)= \frac{1}{4 \pi^2 f_\pi } f_b(t), \nonumber \\
&& f_b(t) =\frac{1}{B ( b -1 , \frac{M_V^2}{a} )} \sum_{n=0}^\infty
\frac{\Gamma(2-b+n)}{\Gamma(n+1) \Gamma(2-b)} \frac1{M_n^2 -t}, \label{eq:reg}
\end{eqnarray}
where $f_b(0)=1$, $f_b(t=-Q^2) \sim (Q^2)^{1-b}$. The same approach
was used in~\cite{RuizArriola:2008sq} to successfully describe the pion
charge form factor, also at large $Q^2$.  Note that $b \le 1.5$
complies to TW~II, while $b>1.5$ violates the bound.  In the right
panel of Fig.~\ref{fig:one} we show the results of several Regge
models: the Veneziano-Dominguez model with the fitted value $b=1.81$
(dashed line), the Veneziano-Dominguez model with the first pole
separated and fixed $b=1.5$ (dotted line), and the subtracted Regge
model (solid line). The details can be found
in~\cite{Arriola:2010aq}. We note that the models differ only at very
large values of $Q^2$, where the present experimental uncertainties
are large.

Hence, it is possible to explain the BaBar data within the Regge
approach, both in the TW~II violating and TW~II conserving
versions.

Along the lines of the discussion around
  Eq.~(\ref{eq:DR}), another adventurous scenario can be proposed, which lies between the one
  mass state case and the infinitely many uniformly spaced squared
  mass states. We can then interpret the constant $c$ for the IVMD ansatz of
  Eq.~(\ref{eq:IVMD}) as corresponding to a high-mass
  state, $M_H$, such that for $M_V^2 \ll Q^2 \ll M_H^2 $ one has a
  $Q^2$-independent behavior.  This can be represented by the simple
  functional form
\begin{eqnarray} 
F_{\pi^0 \gamma \gamma^*} (-Q^2) = \frac{1}{4 \pi^2 f_\pi} \left[  
c \frac{M_V^2}{M_V^2+Q^2} + (1-c) \frac{M_H^2}{M_H^2+Q^2} \right], 
\end{eqnarray}
which resembles the IVMD ansatz of Eq.~(\ref{eq:IVMD}) in
the range $M_V^2 \ll Q^2 \ll M_H^2$. A direct fit to the joint CLEO and BaBar
data, with fixed $M_V=748~{\rm MeV}$, yields $c=0.085(2)$ and the
lowest bound (within the 95\% confidence level) for the high-mass state 
is $M_H \sim 10~{\rm GeV}$.

\section{Photon momentum asymmetry}

The last issue we wish to discuss in this talk is the influence of the photon momentum 
asymmetry parameter, $A$.
In the BaBar kinematic setup $-q_1^2<0.6~{\rm GeV}^2$ and $-q_2^2 >3~{\rm GeV}^2$, 
therefore
\begin{eqnarray} 
|A| = \left | \frac{q_1^2-q_2^2}{q_1^2+q_2^2} \right |  \sim 0.9-0.97 \neq 1.
\end{eqnarray}
This turn out to be significant  for precision fits and the optimum values of their physical parameters.
In particular, the single-state IMVD model becomes
\begin{eqnarray} 
F_{\pi^0 \gamma \gamma^*} (-Q^2) = \frac{1}{4 \pi^2 f_\pi} \left[ 1 -
  c \left(1- \frac{4M_V^4}{4M_V^4+4M_V^2 Q^2+(1-A^2)Q^4} \right) \right ]. 
\end{eqnarray}
The fits to the combined data yield for 
$A=1$, $0.975$, and $0.95$, respectively,
$c=0.986$, $0.978$, $0.974$,
$M_V=748$, $754$, $768~{\rm MeV}$.
In particular, the value of $M_V$ increases toward $m_\rho$ as $A$ decreases.

\section{Conclusions}

These are our main conclusions:

\begin{itemize}

\item The second Terazawa-West bound, following from assumptions for the real part of the pion
  transition form factor, need not be fulfilled in field-theoretic
  approaches. An explicit counter-example with the Georgi-Manohar model with $g_A \neq 1$, 
  where the chiral anomaly is
  fulfilled but $F_{\pi^0 \gamma \gamma^\ast}$ tends to a constant at $t \to
  -\infty$. If this feature holds in QCD, it opens a
  possibility of solving the BaBar puzzle within models incorporating the incomplete vector-meson dominance.

\item The coefficient in the TW~II is large, extending the bound an order of
  magnitude above the data. Thus, even if it holds in the
  absence of polynomial contributions to the real part, it is
  completely ineffective for the presently-available momentum range.

\item An additional constraint on the models in the time-like region
  of momenta follows from the rare \mbox{$Z \to \pi^0 \gamma$} decay.  We use
  this bound in our considerations. For the considered models it is comfortably
  satisfied.

\item The simplest model realizing the incomplete vector-meson
  dominance, including a single vector-meson state, is capable of
  reproducing the data for $F_{\pi^0 \gamma \gamma^\ast}$ 
  in the whole available experimental range, $0 <
  Q^2 < 35~{\rm GeV}^2$.

\item Within the Regge approach with infinitely many radially excited states, 
  the data can be fitted both satisfying or violating
  the TW~II bound. 

\item The precise numerical fits are sensitive to the photon momentum asymmetry
  parameter, $A$, which affect the optimum values of the physical
  parameters, such as the vector meson mass. Since $A$ is not strictly $1$ in the experimental setup, 
  the effects of kinematic cuts should be considered in precision analyses.

\end{itemize}
       

\providecommand{\href}[2]{#2}\begingroup\raggedright\endgroup

\end{document}